# SEMI-SUPERVISED NMF-CNN FOR SOUND EVENT DETECTION

*Teck Kai Chan[1,2], Cheng Siong Chin[1], Ye Li[2]*

Newcastle University Singapore[1], Xylem Water Solution Singapore Pte Ltd[2]

## ABSTRACT

In this paper, a combinative approach using Nonnegative Matrix Factorization (NMF) and Convolutional Neural Network (CNN) is proposed for audio clip Sound Event Detection (SED). The main idea begins with the use of NMF to approximate strong labels for the weakly labeled data. Subsequently, using the approximated strongly labeled data, two different CNNs are trained in a semi-supervised framework where one CNN is used for clip-level prediction and the other for frame-level prediction. Based on this idea, our model can achieve an event-based F1-score of 45.7% on the Detection and Classification of Acoustic Scenes and Events (DCASE) 2020 Challenge Task 4 validation dataset. By ensembling models through averaging the posterior outputs, event-based F1-score can be increased to 48.6%. By comparing with the baseline model, our proposed models outperform the baseline model by over 8%. By testing our models on the DCASE 2020 Challenge Task 4 test set, our models can achieve an event-based F1-score of 44.4% while our ensembled system can achieve an event-based F1-score of 46.3%. Such results have a minimum margin of 7% over the baseline system which demonstrates the robustness of our proposed method on different datasets.

## 1. INTRODUCTION

A Sound Event Detection (SED) system can be described as an intelligent system that is capable of not only detecting the types of sound events present in an audio recording but also returning the temporal location of the detected events. Such a system can be useful in several different domains and as compared to a visual detection system, it can be advantageous in several different aspects. Firstly, a SED system is not affected by the degree of illumination. Secondly, occluded objects do not affect detection accuracy. Thirdly, audio recording requires lesser computational resources as compared to an image or video. Finally, events such as a car horn, can only be detected by sound [1], [2].

However, for a SED system to achieve maximum performance, there may be a need for a large amount of strongly labeled data where the event onset and offset is known with certainty during the model development phase. This can be a limiting factor because such data is usually difficult and time-consuming to collect [3].

As shown in our previous work [4], NMF can be used to approximate strong labels for the weakly labeled data. As a follow-up work, we propose to label the weakly labeled data using NMF in a supervised manner. Using the approximated strongly labeled data, two different CNNs are trained in a semi-supervised framework where one of the models will produce the clip level prediction and the other will produce a frame-level prediction. Based on such framework, our best model can achieve an event-based F1-score of 45.7% on the validation dataset of the DCASE 2020 Challenge Task 4. Using ensemble methods, we can further increase the event-based F1-score to 48.6%. By comparing our models with the baseline model, our models outperformed the baseline model by over 8%. On the other hand, by testing our models on the test dataset of the DCASE 2020 Challenge Task 4, our best model can achieve an event-based F1-score of 44.4% while our ensembled system can achieve an event-based F1-score of 46.3%. Such results have a minimum margin of 7% over the baseline system which demonstrates the robustness of our method on different datasets.

The rest of the paper is organized as follow, Section 2 describes the dataset used, Section 3 describes the proposed methodology. Section 4 provides the results followed by a discussion. Finally, the paper ends with a conclusion.

## 2. DATASET DESCRIPTION

The dataset used is the DCASE Challenge 2020 Task 4 dataset [5]. It consists of ten event labels with different distributions and can be categorized into three different categories: Synthetic Strongly Labeled, Weakly Labeled and Unlabeled. There are a total of 2595 synthetic audio clips, 1578 weakly labeled audio clips, 14412 unlabeled audio clips where each clip has a duration of 10s. The validation set consists of 1168 audio clips where each clip has a similar duration of 10s. Whereas the test set consists of 12566 audio clips where each clip has a duration of 10s or 5 mins.

## 3. PROPOSED METHODOLOGY

### 3.1. Audio Preprocessing and Feature Extraction

As the first step of pre-processing, all audio clips that are longer or shorter than 10s are first truncated or padded to have an equal length of 10s. Processed clips are then resampled at 22,050 Hz, and spectrograms are tabulated using a Fast-Fourier Transform (FFT) window size of 2048 (92 ms) with a hop length of 345 (15.6 ms). Mel-spectrograms are then tabulated using 64 mel filter banks. Based on such a setting, a tabulated mel spectrogram would have a size of 640 by 64, where 640 represents the number of frames, and 64 represents the number of mel bins. Finally, a logarithm operation was applied to obtain the log mel spectrogram, which will be used as model input.

### 3.2. Approximating Strong Labels Using NMF

As shown in our previous work [4], NMF [6] can be used to approximate strong labels for the weakly labeled data. This was done by deriving the activation matrix, $H$, from the mel spectrogram of each audio clip without the use of any dictionary. Each frame was considered to be activated if it exceeds a predefined threshold which in turn suggest the occurrence of an audio event. However, if a clip contained multiple audio events, then those activated frames were assumed to contain all the audio events. Such assumption may not be true for all scenarios, thus, such method can induce noise into the training labels.

In this paper, we propose to approximate strong labels for weakly labeled data in a supervised manner. The first step is to extract the event template from the synthetic audio clips to form a dictionary for different event classes. Since synthetic sound clip can contain multiple events, temporal masking is applied to the mel spectrogram using the given temporal annotations. Templates of each event class are retrieved from the masked mel spectrogram using NMF by allowing the number of component, $r$, to be set as 1. For example, if synthetic clip A has Speech and Cat occurring at frame 1 to 100 and 100 to 110 respectively, all frames from 101 onwards are masked to extract the Speech template followed by masking all frames except frames 100 to 110 to extract the Cat template. Note that for events that are overlapping, we process them in the same manner.

As weakly labeled data possessed the event tags, we apply the corresponding dictionary on the sound clip to derive $H$. Frames that are activated (above a pre-defined threshold) are assumed to contain the event class. For example, if Clip B contains Speech and Dog, we first apply NMF to decompose Clip B using Speech dictionary and with $r$ set as 1 to derive $H$. Frames that are over a threshold are assumed to contain only Speech. A similar procedure is applied to derive the temporal annotation for Dog by using the Dog dictionary instead of Speech dictionary.

After this process, each weakly labeled clip would have an approximated strongly label in a form of a matrix of size 640 by 10 where each column represents an event while the occurrence of an event is indicated as 1 and 0 otherwise for each row.

### 3.3. Semi-supervised Learning

As mentioned in [7], there can be a trade-off in SED performance due to the pooling operation. While the accuracy of clip level detection (also known as audio tagging) can be improved with higher temporal compression (pooling along the time axis), this can result in a degradation of accuracy in frame-level detection.

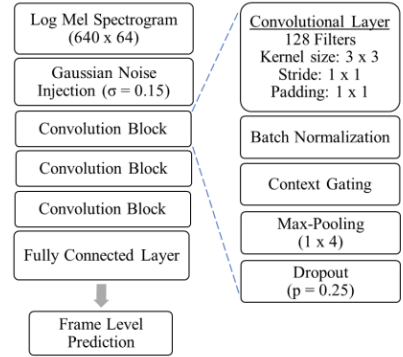

**Fig. 1.** Model for Frame Level Prediction

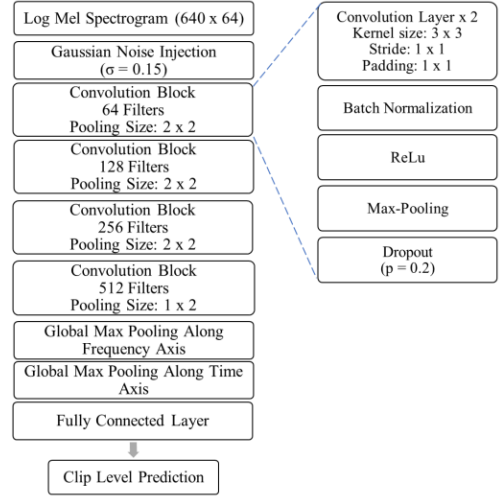

**Fig. 2.** Model for Clip Level Prediction

Therefore, we propose a Shallow Model (SM) with no temporal compression for frame-level prediction and a Deep Model (DM) with temporal compression for clip level prediction. In addition to the difference in pooling size, SM has fewer convolutional layers, adopted context gating [8] as the activation function as opposed to ReLu and has a slightly higher dropout rate. The details of SM and DM can be found in Fig. 1 and Fig. 2 respectively.

Given that $y_f$ and $y_c$ are the frame level and clip level ground truth of an input respectively. The Binary Cross Entropy (BCE) loss between the frame-level prediction of SM, $SM_f$, and $y_f$ can be given as

$$l_f = BCE(SM_f, y_f) \qquad (1)$$

While the BCE loss between the clip level prediction of DM, $DM_c$, and $y_c$ can be given as

$$l_c = BCE(DM_c, y_c) \qquad (2)$$

As mentioned earlier, the accuracy of clip-level detection is better for models with higher temporal compression. We hypothesize that by enforcing the prediction of SM to be consistent with DM, it could produce a better frame-level prediction. As the prediction output of SM is in frame level, we apply a global max pooling on the time axis of $SM_f$ to obtain the clip level prediction, $SM_c$. Instead of using BCE, we propose the use of Mean Square Error (MSE) as the consistency loss function. This was found to be a better consistency loss function as compared to using BCE [9]. However, this loss will only be calculated if DM is confident with its prediction. Thus, the consistency loss is given as

$$l_{con} = \begin{cases} MSE(SM_c, DM_c) > \lambda \\ 0, otherwise \end{cases} \qquad (3)$$

where $\lambda$ represents the confidence level. In addition, we also enforce the consistency of prediction on the unlabeled data. This is also known as semi-supervised learning, which was found to improve the performance and generalization of the model [10], [11]. Given that the clip level prediction from SM and DM on the unlabeled data are $SM_{uc}$ and $DM_{uc}$ respectively, the loss between $SM_{uc}$ and $DM_{uc}$ is represented by $l_{unlabel}$. To regularize the contribution of $l_{unlabel}$, a weighting parameter, $w$ is proposed and can be defined as

$$w = \exp(-5(1-T)^2) \qquad (4)$$

where $T$ is a positive value which represents the training progression. Similarly, $l_{unlabel}$ is calculated only if DM is confident with its prediction. Thus $l_{unlabel}$ is defined as

$$l_{unlabel} = \begin{cases} w \times MSE(SM_{uc}, DM_{uc}) > \lambda \\ 0, otherwise \end{cases} \qquad (5)$$

Based on the calculated losses, parameters of both models will then be updated using Adam [12]. As it was found that the performance of deep NN may benefit from Learning Rate (LR) reset [13], we propose to anneal the LR according to a cosine function and reset it to original LR after a certain number of epochs. The LR at each iteration can be defined as [13]

$$LR_{curr} = LR_{max} + \frac{1}{2}(LR_{max} - LR_{min})\left(1 + \cos\left(\frac{T_{curr}}{T_i}\pi\right)\right) \qquad (6)$$

Where $LR_{max}$ represents the maximum LR and was set as 0.0012. $LR_{min}$ represents the minimum LR which was set as 1e-6. $T_{curr}$ represents the current training iteration and $T_i$ represent the maximum training iterations before a LR reset. The LR will be reset whenever $T_{curr}$ is equal to $T_i$, and at the next iteration, $T_{curr}$ will then be reset to 0 while $T_i$ is multiplied with an integer, $T_{mult}$ which can delay the next restart if $T_{mult}$ is larger than 1.

As defined in Equation 4, $T$ represents the training progression, which directly affects the calculation of $l_{unlabel}$. We proposed to define $T$ as

$$T = \frac{T_{curr}}{T_i} \qquad (7)$$

Thus, $w$ will be reset to 0 whenever the LR is reset.

Finally, we also adopt the concept of transfer learning in our system, where the models will be trained using synthetic data for 5 epochs without the inclusion of $l_{unlabel}$. Only from the 6th epoch onwards, the parameters will be updated using real data and with the inclusion of $l_{unlabel}$

### 3.3. Post Processing

A clip is considered to contain a specific event if the predicted probability from DM is larger than 0.5. Using the identified audio tag, temporal location can be found by locating the activated frames based on the predicted outputs from SM. Outputs from SM are smoothed using iterative median filter [14] with an event-specific window size. Frames are then considered to be activated if they exceeded an event-specific frame threshold. Following [15], neighboring frames are also considered to be activated if they exceeded a lower bound threshold of 0.08. In addition, detected events with a duration of shorter than 0.1s were removed as they are considered as noise. Finally, two similar events are concatenated together if the difference between the first event offset and the second event onset is shorter than 0.2s.

### 4. RESULTS

By applying our proposed framework on the DCASE 2020 Challenge Task 4 validation dataset, Proposed System (PS) 1 can achieve an event-based F1-score of 45.2% by setting $T_i$ as 1 epoch and $T_{mult}$ as 2. As mentioned earlier, models are trained using only synthetic data for the first 5 epochs and only from the 6th epoch onwards, models are trained using weakly labeled and unlabeled data. In our experiment, we also tested a different form of transition where model were trained using both real and synthetic data from the 6[th] epoch onwards. Based on such setting, PS 2 can achieve a slightly higher event based F1-score of 45.7% with a similar $T_i$ and $T_{mult}$ as PS 1. We then averaged the posterior outputs from PS 1 and PS 2 to create an ensemble, PS 3, and the event-based F1-score can be increase to 48.0%. Since each system can have a different optimal median filter window, we then tuned the median filter window of PS 3, and this could further increase the event-based F1-score to 48.6% which is represented as PS 4 in Table 1. Based on such results, our models have a margin of over 8% as compared to the baseline systems [18], [19].

By applying our systems on the DCASE 2020 Challenge Task 4 test dataset, PS 1 has an event-based F1 score of 43.5% while PS 2 has a an event-based F1 score of 44.4%. Similarly, ensembled systems demonstrate better predictive power where PS 3 achieves an event based F1-score of 45.8% while PS 4 achieves an event based F1-score of 46.3%. Such results have a minimum margin of 7% over the baseline system which demonstrates the robustness of our proposed method on different dataset.

We also computed the Polyphonic Sound Detection Score (PSDS) [16] as a secondary measure. Results also shown that our systems have a higher PSDS score which demonstrates the effectiveness of our system.

|  |  | EB F1-Score (%) | PSDS |
|---|---|---|---|
| Validation | PS 1 | 45.2 | 0.630 |
| | PS 2 | 45.7 | 0.635 |
| | PS 3 | 48.0 | **0.652** |
| | PS 4 | **48.6** | 0.649 |
| | Baseline without SS [18] | 34.8 | 0.61 |
| | Baseline with SS [19] | 35.6 | 0.626 |
| Test | PS 1 | 43.5 | 0.503 |
| | PS 2 | 44.4 | 0.522 |
| | PS 3 | 45.8 | **0.543** |
| | PS 4 | **46.3** | 0.534 |
| | Baseline without SS [18] | 34.9 | 0.496 |
| | Baseline with SS [19] | 36.5 | 0.497 |

**Table 1.** Results of Proposed System Against Baseline (EB refers to Event-Based and SS refers to Source Separation)

## 5. DISCUSSION

There are several factors that can affect the accuracy of the system. Firstly, a high value of $\lambda$ is required to prevent a suboptimal solution and was set as 0.9 to ensure that $l_{unlabel}$ will only be calculated based on highly confident prediction.

Secondly, using an event-specific frame threshold, detection accuracy can be raised. However, optimal values differ across systems. Likewise, the median filter window is also dependent on the system trained. These parameters can be found by tuning them against the validation dataset. In our experiments, we found that using a smaller window in the first round of filtering and larger window size in the second round of filtering usually produced higher accuracy.

The post-processing method in [15] began with the joining of similar events before the removal of noise. However, accuracy can be higher if the noise are removed before the concatenation of similar events.

$T_i$ controls how fast LR will reduce from $LR_{max}$ to $LR_{min}$. In our experiment, if $T_i$ is smaller than 5 epochs, $T_{mult}$ must be at least 2 to prevent the large fluctuation of LR throughout the training process. If $T_i$ is larger than 5 epochs, $T_{mult}$ can be set as 1 as the transition of LR from $LR_{max}$ to $LR_{min}$ can be considered slow and steady. Subsequently, we found that it is not a guarantee that a better solution can be found following a LR reset, and there is a possibility that the performance can degrade. However, overall performance of our models do benefit from LR reset.

Subsequently, the use of synthetic data provides several benefits such as speeding up the convergence of the models using transfer learning and also alleviating the complications caused by the presence of noisy strong labels approximated using NMF. As seen in Table 1, accuracy of PS 2 which was trained using a combination of synthetic and real data after the transfer learning phase, was higher than PS 1 which was trained using only real data after the transfer learning phase.

One weaker aspect of our framework is the detection accuracy of Dishes. Using PS 4 as an example, the detection accuracy of Dishes was only marginally above 20% which can be seen in Table 2.

| Event Label | Validation | Test |
|---|---|---|
| Speech | 54.9 | 58.7 |
| Dog | 34.2 | 46.4 |
| Cat | 38.6 | 55.3 |
| Alarm/Bell Ringing | 50.9 | 41.6 |
| Dishes | 22.9 | 20.1 |
| Frying | 54.0 | 50.2 |
| Blender | 47.9 | 55.1 |
| Running Water | 45.5 | 34.3 |
| Vacuum Cleaner | 77.0 | 59.9 |
| Electric Shaver/Toothbrush | 60.5 | 42.2 |

**Table 2.** Classwise Event Based F1-Score of PS 4

## 6. CONCLUSION

In this paper, a combinative approach using NMF and CNN was proposed. Using such framework, the best model can achieve an event-based F1-score of 45.7%, while the ensemble model can achieve an event-based F1-score of 48.6% on the validation dataset of the DCASE 2020 Challenge Task 4. Based on such results, our systems outperform the baseline system by over 8%. On the other hand, the best model can achieve an event-based F1-score of 44.4% while our ensembled system can achieve an event-based F1-score of 46.3% on the test dataset of the DCASE 2020 Challenge Task 4. Such results have a minimum margin of 7% over the baseline system which demonstrates the robustness of our method on different datasets. For our future work, we will investigate the cause of low detection accuracy for Dishes and improve our system in this aspect.